\documentclass[journal]{IEEEtran}
\usepackage{amsmath,amsfonts,amssymb}
\usepackage{algorithmic}
\usepackage{algorithm}
\usepackage{array}
\usepackage[caption=false,font=normalsize,labelfont=sf,textfont=sf]{subfig}
\usepackage{textcomp}
\usepackage{stfloats}
\usepackage{url}
\usepackage{verbatim}
\usepackage{graphicx}
\usepackage{cite}
\usepackage{bm}
\usepackage{bbm}
\usepackage{color}
\newtheorem{lem}{Lemma}
\newtheorem{thm}{Theorem}
\newtheorem{dfn}{Definition}
\newtheorem{ex}{Example}

\begin{document}

\title{Alpha-NML Universal Predictors}

\author{Marco Bondaschi,~\IEEEmembership{Graduate Student Member,~IEEE,}
	and~Michael~Gastpar,~\IEEEmembership{Fellow,~IEEE}%
\thanks{The authors are with the School of Computer and Communication Sciences, \'{E}cole Polytechnique F\'{e}d\'{e}rale de Lausanne, CH-1015 Lausanne, Switzerland (e-mail: \{marco.bondaschi, michael.gastpar\}@epfl.ch).}%
\thanks{This work was presented in part at the 2022 IEEE International Symposium on Information Theory, Espoo, Finland, Jun. 2022.}%
\thanks{This work was supported in part by the Swiss National Science Foundation under Grant 200364.}%
\thanks{Copyright (c) 2024 IEEE. Personal use of this material is permitted.  However, permission to use this material for any other purposes must be obtained from the IEEE by sending a request to pubs-permissions@ieee.org.}}



\maketitle

\begin{abstract}
Inspired by the connection between classical regret measures employed in universal prediction and R\'{e}nyi divergence, we introduce a new class of universal predictors that depend on a real parameter $\alpha\geq 1$. This class interpolates two well-known predictors, the mixture estimators, that include the Laplace and the Krichevsky-Trofimov predictors, and the Normalized Maximum Likelihood (NML) estimator. We point out some advantages of this new class of predictors and study its benefits from two complementary viewpoints: (1) we prove its optimality when the maximal R\'{e}nyi divergence is considered as a regret measure, which can be interpreted operationally as a middle ground between the standard average and worst-case regret measures; (2) we discuss how it can be employed when NML is not a viable option, as an alternative to other predictors such as Luckiness NML. Finally, we apply the $\alpha$-NML predictor to the class of discrete memoryless sources (DMS), where we derive simple formulas to compute the predictor and analyze its asymptotic performance in terms of worst-case regret.
\end{abstract}

\begin{IEEEkeywords}
Universal prediction, universal compression, Normalized Maximum Likelihood, Sibson's mutual information, R\'{e}nyi capacity, redundancy-capacity theorem.
\end{IEEEkeywords}

\section{Introduction}
\label{Introduction}
Prediction refers to the general problem of estimating the next symbols of a sequence given its past, and evaluating the confidence of such an estimate. This problem appears in a large number of research areas, such as information theory, statistical decision theory, finance, and machine learning. Some knowledge about the probability distribution that models the sequence one wishes to predict is clearly helpful. Unfortunately, in many practical applications such knowledge is missing. If this is the case, then one may wish to say something about the future of the sequence when the true model of the source that is producing the symbols is \emph{any} of the models belonging to a certain class. This problem usually goes under the name of \emph{universal prediction} \cite{merhav1}. It has applications in a wide range of areas, such as compression \cite{ziv1, willems1}, gambling \cite{xie1} and machine learning \cite{fogel1,rosas1}.

The formal statement of universal prediction that we consider in this work is the following. For any $n\geq 1$, we assume that a sequence $x^{n-1} = (x_1,x_2,\dots,x_{n-1})$ of $n-1$ symbols from a given (possibly infinite) alphabet $\mathcal{X}$ has been generated by some unknown (random or deterministic) source. Suppose that we design a predictor that, given the past symbols of the sequence, returns some numerical prediction about the next symbol $x_n$. This prediction may be an estimation $\hat{x}_n$ of next symbol itself, or it may also be something more informative, such as an estimation of the probability distribution of the next symbol. The latter case carries the additional information of the \emph{confidence} associated to the estimation, in terms of how probable our best guess on the next symbol is.

In order to evaluate the quality of the prediction, one uses a so-called \emph{loss function} $\ell$ that maps the pair formed by the prediction and the actual symbol $x_n$, to a real number. In this paper, we follow the classical perspective where the predictor assigns probabilities to the possible values of the next outcome $x_n$ \cite{merhav1, cesa1}. In such a case, one usually chooses as a loss function some value that is inversely proportional to the estimated probability of $x_n$. The reason for such a choice is that, if the source generates frequently symbols to which our predictor assigned a low probability, then the measured loss is high, signaling that our predictor is bad; on the contrary, if the source generates symbols to which the predictor assigned high probability, then the loss is small.

A very popular choice, mainly due to its connection with universal compression, is the \emph{logarithmic loss}. If $\hat{p}(\cdot|x_1^{n-1})$ is the probability distribution on the next symbol estimated by the predictor, then the associated logarithmic loss is defined as
\begin{equation}
\ell(\hat{p},x_n) \triangleq \log\frac{1}{\hat{p}(x_n|x^{n-1})}.
\end{equation}
If the quality of the predictor is measured on more than one symbol, for example the whole sequence $x^n$, then one can take as a performance measure the \emph{cumulative loss} $L$, which is the sum of the losses of the $n$ symbols. In the case of the logarithmic loss, one has
\begin{align}
L(\hat{p},x^n) &\triangleq \sum_{i=1}^n \ell(\hat{p}, x_i) \\
	&= \sum_{i=1}^n \log\frac{1}{\hat{p}(x_i|x^{i-1})} \\
	&= \log \frac{1}{\hat{p}(x^n)}
\end{align}
where $\hat{p}(x^n) \triangleq \prod_{i=1}^n \hat{p}(x_i|x^{i-1})$ can be defined as the joint estimated probability of the entire sequence $x^n$. In the remainder of the paper, we take our loss function to be the logarithmic loss.

Let us now consider a given class of distributions $\mathcal{P} = \{p_{\theta}: \theta\in\Theta\}$ indexed by a parameter set $\Theta$, and let us assume that the actual source belongs to this class, or, less strictly speaking, that this class is the one we want to compare our predictor to. Usually, $\Theta$ is a subset of $\mathbb{R}^d$ for some $d\geq 1$, and $\theta\in\Theta$ is the parameter vector of some parametric family, e.g., discrete memoryless sources, Markov sources of order $k$, auto-regressive sources, a certain exponential family, etc. When building a predictor for sequences of symbols, one needs a metric or a criterion that measures the quality of the predictor by taking into consideration the different possible sequences $x^n$, as well as the possible sources of the class $\mathcal{P}$. 
To construct such a measure, one usually starts from the difference between the logarithmic loss of the predictor $\hat{p}$ and that of a distribution $p_{\theta}$ in $\mathcal{P}$, that is,
\begin{align}
R(\hat{p},p_{\theta},x^n) &\triangleq \log \frac{1}{\hat{p}(x^n)} - \log \frac{1}{p_{\theta}(x^n)} \\
	&= \log\frac{p_{\theta}(x^n)}{\hat{p}(x^n)},
\end{align}
which is usually called \emph{regret}. Two regret measures that are generally employed to assess the quality of a predictor are the \emph{average regret}
\begin{align}
R_{\rm av}(\hat{p}) &\triangleq \sup_{\theta\in\Theta}\mathbb{E}_{\theta}\left[R(\hat{p},p_{\theta},X^n)\right] \\ \label{AvR}
	&= \sup_{\theta\in\Theta}\int_{\mathcal{X}^n} p_{\theta}(x^n) \log\frac{p_{\theta}(x^n)}{\hat{p}(x^n)} \,dx^n \\
	&= \sup_{\theta\in\Theta} D(p_{\theta}\|\hat{p}), \label{RavDiv}
\end{align}
and the \emph{worst-case regret}
\begin{align}
R_{\max}(\hat{p}) &\triangleq \sup_{\theta\in\Theta}\sup_{x^n \in\mathcal{X}^n} R(\hat{p},p_{\theta},x^n) \\ \label{WCR}
	&= \sup_{\theta\in\Theta}\sup_{x^n\in\mathcal{X}^n} \log\frac{p_{\theta}(x^n)}{\hat{p}(x^n)}\\
	&= \sup_{\theta\in\Theta} D_{\infty}(p_{\theta}\|\hat{p}) \label{RmaxDiv}
\end{align}
where $D_{\infty}(p_{\theta}\| \hat{p})$ is the R\'{e}nyi divergence of order infinity.
The maximization over all parameters in $\Theta$ that appears in the considered definitions of regret comes from the fact that, in the universal prediction setting, one generally considers the case where no prior knowledge on the parameters is available, that is, no source in $\mathcal{P}$ is considered a better candidate to be the true one in advance.


It is well known \cite{shtarkov1} that the predictor that minimizes the worst-case regret is the Normalized Maximum Likelihood (NML) estimator, whenever it exists. Its formula is
\begin{equation}
\label{NML}
\hat{p}_{\rm NML}(x^n) = \frac{\sup_{\theta\in\Theta} p_{\theta}(x^n)}{\int_{\mathcal{X}^n}\sup_{\theta\in\Theta} p_{\theta}(x^n)\,dx^n}.
\end{equation}
Even if it has a nice closed-form expression, in general the NML has several disadvantages, including the fact that it may not exist since the integral in the denominator in \eqref{NML} may not converge, and the necessity of computing a maximization over the parameter space $\Theta$.

These limitations led researchers to look for good alternatives to the NML predictor. For the class of discrete memoryless sources over a finite alphabet $\mathcal{X}=\{1,2,\dots,m\}$, such an alternative is the Krichevsky-Trofimov estimator \cite{krichevsky1}, which assigns as a probability for the next symbol $k\in\{1,2,\dots,m\}$ a value proportional to
\begin{equation}
\label{KTCondEx}
\hat{p}_{\rm KT}(k|x^{n-1}) \propto n_k + \frac{1}{2}\,,
\end{equation}
where $n_k$ is the number of $k$'s in the past sequence $x^{n-1}$.

As opposed to NML, the KT predictor is not affected by the disadvantages listed above. Furthermore, it turns out that it achieves, for the class of discrete memoryless sources, the same asymptotic regret, up to a constant term, as the NML when $n\to\infty$ \cite{xie1}. However, no similar results are proved for other classes of distributions, and also, the NML estimator performs better in general when $n$ is finite. For these reasons, the search for alternative predictors that have fewer drawbacks (e.g., broader applicability) with respect to the NML estimator is still important.

\subsection{Contributions}
The contribution of this paper is the introduction of a class of predictors inspired by Sibson's $\alpha$-mutual information and the connection between regret measures used in universal prediction and R\'{e}nyi divergence, that we term $\alpha$-NML predictors. This class is parametrized by $\alpha \geq 1$ and its definition depends on the choice of a prior probability distribution $w$ over the parameter space $\Theta$:
\begin{equation}
\hat{p}_{\alpha}(x^n) \triangleq \frac{\left\{\int_{\Theta}w(\theta)\,p_{\theta}^{\alpha}(x^n)\,d\theta\right\}^{1/\alpha}}{\int_{\mathcal{X}^n} \left\{\int_{\Theta}w(\theta)\,p_{\theta}^{\alpha}(\bar{x}^n)\,d\theta\right\}^{1/\alpha}\,d\bar{x}^n}\,.
\end{equation}
As an example, for DMS this class interpolates between the KT estimator and the NML. For $\alpha = 1$, our predictor gives the same probability estimation \eqref{KTCondEx} as the KT predictor. For $\alpha = 2$, it assigns a probability that is proportional to
\begin{equation}
\label{Alpha2Cond}
\hat{p}_{\alpha=2}(k|x^{n-1}) \propto \sqrt{\left(n_k + \frac{1}{4}\right)\left(n_k + \frac{3}{4}\right)}\,.
\end{equation}
Finally, when $\alpha\to\infty$, we retrieve the classical NML formula. The general conditional formula as a function of $\alpha$ is given in Equation \eqref{CondProbFinal}.

In the paper, we study the $\alpha$-NML predictors from two complementary perspectives. The first one is to investigate its performance when R\'{e}nyi divergence is used as a regret measure. We call such regret measures $\alpha$-regret; they can be interpreted operationally as a middle ground between the classical average regret and the worst-case regret, which are widely employed in universal prediction. In fact, both the average regret and the worst-case regret can be written as a maximization of a R\'{e}nyi divergence -- see \eqref{RavDiv} and \eqref{RmaxDiv}. If the maximization of a R\'{e}nyi divergence of any order $\alpha$ between these two extreme cases is taken as a regret measure, then it turns out that $\alpha$-NML is the optimal predictor, provided that the proper prior distribution on the parameter space $\Theta$ is chosen.

The second perspective is to look at the $\alpha$-NML predictors as an alternative to NML, when the latter cannot be used because, e.g., it does not exist. In fact, $\alpha$-NML converges to NML when $\alpha \to\infty$. However, if a finite $\alpha$ is chosen, it can be shown that $\alpha$-NML exists for some classes of distributions for which the NML does not. As a further example, we analyze $\alpha$-NML's performance in terms of worst-case regret, which is the measure under which the NML is optimal, and we investigate how much we pay in terms of regret with respect to NML, and how much we gain with respect to the KT predictor, as a function of the parameter $\alpha$, when the class of discrete memoryless sources is considered. For the binary alphabet case, the performance improvement of the new predictor is illustrated numerically in Figure 1.


\subsection{Related work}
The worst-case regret and the Normalized Maximum Likelihood predictor were first studied in \cite{shtarkov1}. The Krichevsky-Trofimov predictor was introduced in \cite{krichevsky1} for binary sources, and it was generalized to general finite alphabets in \cite{xie1}, where its asymptotic worst-case regret is also analyzed. A summary of the properties of Sibson's $\alpha$-mutual information can be found in \cite{verdu1,esposito2024sibsonsalphamutualinformationvariational}. The problem of maximizing Sibson's mutual information is studied in \cite{cai1}, where a result similar to Theorem \ref{MaxIAlpha} of this paper is derived by different means. In \cite{yagli1}, a different regret measure, also based on the R\'{e}nyi divergence, is introduced. In the same paper, the authors show that, in the case of discrete memoryless sources with finite alphabet, their regret measure is equivalent to the $\alpha$-regret defined in \eqref{LambdaRegret}. For this particular case, the authors also derive the asymptotical value of the regret as the sequence length goes to infinity. This result was used in the proof of Theorem \ref{RAlphaNMLAsymptotics} here. In \cite{drmota1}, the authors study the minimax regret with the additional constraint that the point-wise regret be an integer, which naturally arises from a universal compression perspective, where codeword lengths are considered. In \cite{bondaschi1, fogel2}, batch universal prediction is studied, where batches of training data are available to the model to aid prediction.

\subsection{Overview}
The remainder of the paper is organized as follows. In Section \ref{ARegret} we introduce $\alpha$-regret, which is defined in terms of R\'{e}nyi divergence and described from an operational point of view. In Section \ref{ANML}, we introduce the class of $\alpha$-NML predictors and we prove their optimality under $\alpha$-regret. In Section \ref{Alternative}, we discuss the advantages of using $\alpha$-NML as an alternative to NML, when the latter cannot be used, and we compare it to other alternatives to NML such as Luckiness NML. In Section \ref{ANMLDMS}, we apply $\alpha$-NML to the parametric family of discrete memoryless sources (DMS), deriving some simple closed-form formulae to compute the probabilities estimated by the predictor, and studying its performance in terms of worst-case regret, discussing how much we pay by using $\alpha$-NML instead of NML in this setting.

\subsection{Notation}
We use upper case letters $X$ to denote random variables and lower case letters $x$ to denote their realizations. A sequence of length $n$ is denoted by a superscript $X^n.$
Probability distributions are denoted by $p_\theta(x),$ where $\theta \in \Theta$ is a parameter.
The R\'{e}nyi divergence of order $\alpha$ between two probability distributions $p_{\theta}$ and $\hat{p}$ on some alphabet $\mathcal{X}^n$ is denoted as
\begin{equation}
\label{AlphaDivergence}
 D_{\alpha}(p_{\theta}\|\hat{p}) \triangleq \frac{1}{\alpha-1} \log \int_{\mathcal{X}^n} p_{\theta}(x^n) \left(\frac{p_{\theta}(x^n)}{\hat{p}(x^n)}\right)^{\alpha -1} \,dx^n .
\end{equation}
Sibson's $\alpha$-mutual information~\cite{Sibson:69} (see also~\cite{verdu1,esposito2024sibsonsalphamutualinformationvariational}) between two random variables $(X,Y) \sim p_X\times p_{Y|X}$ on $\mathcal{X}\times\mathcal{Y}$ is denoted as
\begin{equation}
\label{IAlphaDef}
I_{\alpha}(X, Y) \triangleq \frac{\alpha}{\alpha - 1} \log \int_{\mathcal{Y}} \left\{\int_{\mathcal{X}} p_X(x)\,p_{Y|X}^{\alpha}(y|x)\,dx\right\}^{1/\alpha}\,dy .
\end{equation}
In the limit $\alpha\to\infty$, the two previous definitions become
\begin{equation}
\label{DInf}
 D_{\infty}(p_{\theta}\|\hat{p}) = \sup_{x^n} \log\frac{p_{\theta}(x^n)}{\hat{p}(x^n)}
\end{equation}
and
\begin{equation}
\label{IInf}
I_{\infty}(X, Y) = \log \int_{\mathcal{Y}} \sup_{x\in\mathrm{supp}(X)} p_{Y|X}(y|x)\,dy
\end{equation}
where $\mathrm{supp}(X)$ is the support of the random variable $X$.

\section{$\alpha$-regret}
\label{ARegret}
The average regret \eqref{AvR} and the worst-case regret \eqref{WCR} are the two most widely employed regret measures in universal prediction. One can interpret them operationally as follows. Suppose that a source $S_{\theta}$ generates sequences of $n$ symbols from an alphabet $\mathcal{X}$ according to a distribution $p_{\theta}$ on $\mathcal{X}^n$, for some parameter $\theta \in \Theta$. For a predictor $\hat{p}$, the regret, i.e., the difference in logarithmic loss, for a given sequence $x^n\in\mathcal{X}^n$ is $R_{\theta}(\hat{p},x^n) = \log\frac{1}{\hat{p}(x^n)} - \log\frac{1}{p_{\theta}(x^n)} = \log\frac{p_{\theta}(x^n)}{\hat{p}(x^n)}$. One then has the choice of how much weight to give to each sequence $x^n$ with respect to the others. The two approaches that we discussed before are the following.
\begin{itemize}
\item Giving weight to each sequence $x^n$ according to its probability. With this choice we obtain the average regret \eqref{AvR}:
\begin{align}
\label{RavTheta}
R_{\mathrm{av},\theta}(\hat{p}) &\triangleq \mathbb{E}_{\theta}\left[\log\frac{p_{\theta}(X^n)}{\hat{p}(X^n)}\right] \\
&= \int_{\mathcal{X}^n} p_{\theta}(x^n)\log\frac{p_{\theta}(x^n)}{\hat{p}(x^n)}\, dx^n = D(p_{\theta} \| \hat{p}).
\end{align}
\item Considering only the sequence $x^n$ with the highest regret. In this case we get the worst-case regret \eqref{WCR}:
\begin{equation}
\label{RmaxTheta}
R_{\max,\theta}(\hat{p}) \triangleq \sup_{x^n} \log\frac{p_{\theta}(x^n)}{\hat{p}(x^n)} = D_{\infty}(p_{\theta},\hat{p}).
\end{equation}
\end{itemize}

Essentially, in the first case we are measuring the goodness of the predictor $\hat{p}$ in terms of its average regret when the sequences are generated by $S_{\theta}$ according to $p_{\theta}$, while in the second case we only consider the regret for the worst possible sequence. The two measures \eqref{RavTheta} and \eqref{RmaxTheta} can be recognized as two extreme cases. The former weighs the sequences only according to their probability, without taking into account the amount of regret each sequence carries. Instead, the latter puts all the weight on the worst sequence, without considering how probable it is for this sequence to actually occur. A natural interpolation between these two cases is given by the following $\alpha$-regret, which is equal to the R\'{e}nyi divergence of order $\alpha$.
\begin{dfn}
    The $\alpha$-regret of a predictor $\hat{p}$ with respect to a distribution $p_{\theta}$ is equal to
    \begin{align}
    R_{\alpha,\theta}(\hat{p}) &\triangleq D_{\alpha}(p_{\theta}\|\hat{p}) \\
    &= \frac{1}{\alpha-1} \log \int_{\mathcal{X}^n} p_{\theta}(x^n) \left(\frac{p_{\theta}(x^n)}{\hat{p}(x^n)}\right)^{\alpha -1} \,dx^n . \label{AlphaRegret}
    \end{align}
\end{dfn}

Note that $R_{\alpha=1,\theta}(\hat{p}) = R_{\rm{av},\theta}(\hat{p})$ and $R_{\alpha=\infty,\theta}(\hat{p})= R_{\max,\theta}(\hat{p})$. This regret measure takes into account all the sequences according to their probability, and at the same time gives some additional penalty to sequences with large regret. In fact, setting $\lambda = \alpha - 1$, one can rewrite this measure as
\begin{align}
\label{RLambdaTheta}
R_{1+\lambda,\theta}(\hat{p}) &\triangleq D_{1+\lambda}(p_{\theta}\|\hat{p}) \\
&= \frac{1}{\lambda} \log\int_{\mathcal{X}^n}p_{\theta}(x^n)\exp\left(\lambda\log\frac{p_{\theta}(x^n)}{\hat{p}(x^n)}\right) \,dx^n
\end{align}
which is an exponential average of the regrets: the larger the value assigned to the parameter $\lambda$, the more importance is given to the regret of each sequence $x^n$ in determining the weighting. Finally, maximizing the regret measures \eqref{RavTheta} and \eqref{RmaxTheta} over all possible sources gives back the definitions of $R_{\rm av}(\hat{p})$ and $R_{\max}(\hat{p})$ in Equations \eqref{AvR} and \eqref{WCR}. Doing the same for $\alpha$-regret gives
\begin{align}
\label{LambdaRegret}
R_{\alpha}(\hat{p}) &\triangleq \sup_{\theta\in\Theta} D_{\alpha}(p_{\theta}\|\hat{p}) \\
&= \sup_{\theta\in\Theta} \frac{1}{\alpha-1} \log \int_{\mathcal{X}^n} p_{\theta}(x^n) \left(\frac{p_{\theta}(x^n)}{\hat{p}(x^n)}\right)^{\alpha -1} \,dx^n
\end{align}
where the parameter $\alpha$ in \eqref{LambdaRegret} and $\lambda$ in \eqref{RLambdaTheta} are related by the equation $\alpha = 1+\lambda$.

It is worth noting that the exponential dependency of $\alpha$-regret has a similar flavor to the codeword length measure that Campbell studied in \cite{campbell1} for compression. In fact, it is well known that there is a strong connection between prediction and compression when the logarithmic loss is used. A classical variable-length coding problem is to find a uniquely decodable code for the symbols in $\mathcal{X}^n$ that minimizes the expected length
\begin{equation}
\label{ExpCodewordLength}
L_{\mathrm{av},\theta} \triangleq \mathbb{E}_{\theta}[\ell(X^n)] = \int_{\mathcal{X}^n} p_{\theta}(x^n)\, \ell(x^n) \,dx^n
\end{equation}
where $\ell(x^n)$ is the length of the codeword associated to the sequence $x^n$, and $X^n$ is distributed according to a given $p_{\theta}$. It is well known that the code that minimizes this quantity is any code that associates to the sequence $x^n$ a codeword with length $\ell(x^n) = \log\frac{1}{p_{\theta}(x^n)}$. If the code is instead constructed using a different distribution $\hat{p}(x^n)$, assigning to $x^n$ a codeword of length $\hat{\ell}(x^n) = \log\frac{1}{\hat{p}}(x^n)$, the penalty payed in number of bits would be exactly the regret $R_{\theta}(\hat{p},x^n) = \log\frac{p_{\theta}(x^n)}{\hat{p}(x^n)}$ seen before. In \cite{campbell1}, Campbell was looking for the optimal code that minimizes an alternative measure to \eqref{ExpCodewordLength} in which an exponential dependency on the codeword lengths $\ell(x^n)$ is introduced, instead of the usual expected codeword length. This is different than the setting we consider here, from which $\alpha$-regret emerges. In fact, from a compression perspective, in this work we are still considering the optimal code with respect to the classical expected codeword length, and the exponential dependency is on the difference in bits (i.e., the regret) that the designed code uses with respect to the optimal one (in the usual expected codeword length sense).

\section{$\alpha$-NML predictors}
\label{ANML}
\subsection{Definition}
With the following definition, we introduce a novel class of universal predictors, that we call $\alpha$-NML predictors.
\begin{dfn}\label{def:main:alphanml}
Let $\mathcal{P} = \{p_{\theta} : \theta\in\Theta\}$ be a parametric class of distributions on an alphabet $\mathcal{X}^n$. Let $w$ be any probability distribution on $\Theta$. For any $\alpha \geq 1$, the $\alpha$-NML predictor is defined as
\begin{equation}
\label{AlphaNML}
\hat{p}_{\alpha}(x^n) \triangleq \frac{\left\{\int_{\Theta}w(\theta)\,p_{\theta}^{\alpha}(x^n)\,d\theta\right\}^{1/\alpha}}{\int_{\mathcal{X}^n} \left\{\int_{\Theta}w(\theta)\,p_{\theta}^{\alpha}(\bar{x}^n)\,d\theta\right\}^{1/\alpha}\,d\bar{x}^n}\,.
\end{equation}
\end{dfn}

Note that the definition of $\alpha$-NML also depends on the class of distributions $\mathcal{P}$ and on the prior distribution $w$ on the parameter space $\Theta$. We omit this dependence to ease the notation, since it will be made clear from the context.
The $\alpha$-NML class is a continuous interpolation between the NML predictor and another very popular class of predictors. In fact, taking $\alpha=1$ gives
\begin{equation}
\label{MixtureDef}
\hat{p}_1(x^n) = \int_{\Theta} w(\theta)\,p_{\theta}(x^n)\,d\theta
\end{equation}
which is the well-known class of \emph{mixture estimators}. Taking instead the limit $\alpha\to\infty$ gives
\begin{equation}
\hat{p}_{\infty}(x^n) = \frac{\sup_{\theta\in\mathrm{supp}(w)} p_{\theta}(x^n)}{\int_{\mathcal{X}^n}\sup_{\theta\in\mathrm{supp}(w)} p_{\theta}(x^n)\,dx^n}.
\end{equation}
When the support of the prior distribution $w$ is the entire parameter space $\Theta$, the $\alpha$-NML for $\alpha=\infty$ is precisely the NML predictor defined in \eqref{NML}. It is important to highlight the role of Sibson's $\alpha$-mutual information in how $\alpha$-NML is defined. Note that the denominator of the NML predictor in \eqref{NML} is ${\exp I_{\infty}(\phi, X^n)}$. The $\alpha$-NML definition ideally follows by replacing Sibson's mutual information of order infinity with any other order $\alpha \geq 1$ in the predictor's denominator. In fact, the denominator in \eqref{AlphaNML} is precisely $\exp \frac{\alpha-1}{\alpha}I_{\alpha}(\phi, X^n)$, where Sibson's mutual information $I_{\alpha}$ is defined in \eqref{IAlphaDef}.

Definition~\ref{def:main:alphanml} begs two fundamental questions. Namely, {\it (i)} the conditions on $\mathcal{P}, w(\theta)$ and $\alpha$ under which the $\alpha$-NML predictor exists, and {\it (ii)} the criteria for choosing $w(\theta)$ and $\alpha.$
We defer the discussion on this topic to Section \ref{Alternative}, where we compare existence conditions for the $\alpha$-NML to those for the NML, in order to investigate the use of $\alpha$-NML as an alternative to NML, when the latter does not converge.

\subsection{Optimality with respect to $\alpha$-regret}
It is a well-known fact (see, e.g., \cite[Thm. 37]{erven1}) that the NML defined in \eqref{NML}, whenever it exists, achieves the minimum worst-case regret \eqref{WCR}, which is equal to
\begin{align}
R_{\max}(\hat{p}_{\rm NML}) &= \min_{\hat{p}} R_{\max}(\hat{p}) = I_{\infty}(\phi, X^n) \\
&= \log \int_{\mathcal{X}} \sup_{\theta\in\Theta} p_{\theta}(x^n)\, dx^n. \label{NMLWCRegret}
\end{align}
Furthermore, it is known and it has been proved several times in different contexts (see \cite{merhav1} and references therein) that, under certain conditions, a mixture predictor of the form \eqref{MixtureDef} is optimal under average regret as defined in \eqref{AvR}, for a proper choice of the prior distribution $w(\theta)$. Since $\alpha$-regret is an interpolation between average and worst-case regret, and, furthermore, the $\alpha$-NML predictor is an interpolation between the mixture predictor and the NML, it is natural to ask whether the $\alpha$-NML is actually optimal under $\alpha$-regret. The following theorem shows that, under suitable conditions, this is indeed the case, when the proper prior distribution $w(\theta)$ is chosen. A similar result with a different proof is shown in \cite{cai1}, where the problem of the maximization of Sibson's $\alpha$-mutual information is investigated.

\begin{thm}
\label{MaxIAlpha}
Let $\mathcal{X}$ be a (possibly infinite) alphabet set, and let $\mathcal{P} = \{p_{\theta} : \theta\in\Theta\}$ be a parametric class of distributions on $\mathcal{X}^n$. Let $(\phi, X^n) \sim w \times p_{\phi}(X^n)$ for some prior distribution $w$ on $\Theta$, and suppose that there exists a probability distribution $w^*$ on $\Theta$ such that, for $(\phi^*, X^n) \sim w^* \times p_{\phi^*}(X^n)$
\begin{equation}
I_{\alpha}(\phi^*,X^n) = \sup_{w: \phi\sim w} I_{\alpha}(\phi, X^n).
\end{equation}
Then, the $\alpha$-NML defined in \eqref{AlphaNML} with prior $w^*$, i.e.,
\begin{equation}
\label{PAlphaStar}
\hat{p}_{\alpha}(x^n) = \frac{\left\{\int_{\Theta} w^*(\theta)p_{\theta}^{\alpha}(x^n)\,d\theta\right\}^{1/\alpha}}{\sum_{x^n} \left\{\int_{\Theta} w^*(\theta)p_{\theta}^{\alpha}(x^n)\,d\theta\right\}^{1/\alpha}},
\end{equation}
minimizes $R_{\alpha}(\hat{p})$ over all probability distributions on $\mathcal{X}^n$. Furthermore, the minimal $\alpha$-regret is equal to
\begin{equation}
\label{MinAlphaRegret}
\min_{\hat{p}} R_{\alpha}(\hat{p}) = I_{\alpha}(\phi^*,X^n).
\end{equation}
\end{thm}
\begin{IEEEproof}
The case for $\alpha=1$ is well known and was first proved by Gallager in \cite{gallager1}. We prove the theorem for $\alpha > 1$, following an idea similar to Gallager's. Let $C_{\alpha} \triangleq  \sup_{w} I_{\alpha}(\phi, X^n)$. We want to prove that
\begin{equation}
D_{\alpha}(p_{\theta}\|\hat{p}_{\alpha}) \leq C_{\alpha}
\end{equation}
for every $\theta\in\Theta$. By contradiction, suppose that there exists $\bar{\theta}\in\Theta$ such that,
\begin{equation}
D_{\alpha}(p_{\bar{\theta}}\|\hat{p}_{\alpha}) > C_{\alpha}.
\end{equation}
For any $0\leq t \leq 1$, define the probability distribution
\begin{equation}
w^*_{\bar{\theta},t} = (1-t)w^* + t \delta_{\bar{\theta}}
\end{equation}
where $\delta_{\bar{\theta}}$ is the singular distribution centered on $\bar{\theta}$. Then, we have
\begin{align}
f(t) &\triangleq (\alpha-1)I_{\alpha}(\phi, X^n)\big|_{\phi\sim w^*_{\bar{\theta}}} \\
&= \alpha\log \int_{\mathcal{X}^n} \Big\{t p_{\bar{\theta}}^{\alpha}(x^n) \notag \\
&\hspace{4em}+ (1-t)\int_{\Theta} w^*(\theta)p_{\theta}^{\alpha}(x^n)\,d\theta\Big\}^{1/\alpha} dx^n.
\end{align}
By the assumption that $w^*$ is the maximizer of $I_{\alpha}(\phi, X^n)$, $f(t)$ is maximized at $t=0$.
Taking the derivative of $f(t)$ with respect to $t$ gives
\begin{align}
f'(t) &= \frac{1}{\int_{\mathcal{X}^n}\left\{t p_{\bar{\theta}}^{\alpha}(x^n)+(1-t)\int_{\Theta}w^*(\theta)p_{\theta}^{\alpha}(x^n)d\theta\right\}^{1/\alpha} dx^n} \notag\\
&\hspace{4em}\cdot\int_{\mathcal{X}^n} \left(p_{\bar{\theta}}^{\alpha}(x^n) - \int_{\Theta}w^*(\theta)p_{\theta}^{\alpha}(x^n)d\theta\right) \notag\\
&\hspace{10em}\left\{\int_{\Theta}w^*(\theta)p_{\theta}^{\alpha}(x^n)d\theta\right\}^{\frac{1-\alpha}{\alpha}}dx^n
\end{align}
Evaluating this derivative in $t=0$ gives
\begin{align}
f'(0) &= \frac{\int_{\mathcal{X}^n} p_{\bar{\theta}}^{\alpha}(x^n) \left\{\int_{\Theta}w^*(\theta)p_{\theta}^{\alpha}(x^n)d\theta\right\}^{\frac{1-\alpha}{\alpha}}dx^n}{\int_{\mathcal{X}^n}\left\{\int_{\Theta}w^*(\theta)p_{\theta}^{\alpha}(x^n)d\theta\right\}^{1/\alpha}dx^n}-1 \\
	&= \exp\big\{(\alpha-1)(D_{\alpha}(p_{\theta}\|\hat{p}_{\alpha}) - C_{\alpha})\big\} - 1 > 0.
\end{align}
This contradicts the fact that $f(t)$ is maximized at $t=0$, so we proved that $D_{\alpha}(p_{\theta}\|\hat{p}_{\alpha}) \leq C_{\alpha}$ for every $\theta$. Hence,
\begin{equation}
R_{\alpha}(\hat{p}_{\alpha}) = \max_{\theta} D_{\alpha}(p_{\theta}\|\hat{p}_{\alpha}) \leq C_{\alpha}.
\end{equation}
However, it is known \cite{yagli1} that $\min_{\hat{p}} R_{\alpha}(\hat{p}) = C_{\alpha}$, which proves that $\hat{p}_{\alpha}$ with prior $w^*$ is indeed a minimizer of $R_{\alpha}(\hat{p})$. Substituting \eqref{PAlphaStar} into the definition of $\alpha$-regret \eqref{AlphaRegret} gives precisely $I_{\alpha}(\phi^*, X^n)$, proving \eqref{MinAlphaRegret}.
\end{IEEEproof}


\section{$\alpha$-NML as an alternative to NML}
\label{Alternative}
An important feature of the class of $\alpha$-NML predictors is that these predictors are able to solve some of the problems that afflict the classical NML. First of all, $\alpha$-NML predictors do not require any maximization over the parameter space $\Theta$. The maximization is in fact replaced by a weighted average of the distributions $p_{\theta}$ to the power of $\alpha$. Furthermore, by choosing carefully the prior $w$ and the parameter $\alpha$, one is able to control the convergence of the integral at the denominator of \eqref{AlphaNML}. In this sense, the role of the prior $w$ is similar to that of the luckiness function that appears in the definition of Luckiness NML \cite{grunwald1, grunwald2}, an alternative predictor that was introduced in the literature to overcome the convergence problem of the NML estimator. It is defined as
\begin{equation}
\label{LNML}
\hat{p}_{\rm LNML}(x^n) = \frac{\sup_{\theta\in\Theta}\pi(\theta)p_{\theta}(x^n)}{\sum_{\bar{x}^n\in\mathcal{X}^n}\sup_{\theta\in\Theta}\pi(\theta)p_{\theta}(\bar{x}^n)}.
\end{equation}
where $\pi$ is the luckiness function, a probability distribution on $\Theta$ which models how confident one is that a given $\theta \in \Theta$ is the true parameter of the source. Luckiness NML is the predictor that minimizes a regret measure related to the worst-case regret \eqref{WCR}, the \emph{worst-case luckiness regret}, which is defined as
\begin{equation}
\label{WCLRegret}
R_{\max}(\pi, \hat{p}) = \max_{\theta\in\Theta} \max_{x^n} \frac{\pi(\theta)p_{\theta}(x^n)}{\hat{p}(x^n)}.
\end{equation}
However, $\alpha$-NML has three additional advantages with respect to Luckiness NML, as an alternative predictor to NML:
\begin{itemize}
    \item it has a simple interpretation in terms of R\'{e}nyi divergence, and it is provably optimal under $\alpha$-regret measures;
    \item provided that $\mathrm{supp}(w) = \Theta$, the $\alpha$-NML converges to NML (if it exists) as $\alpha \to\infty$, while this is not the case for Luckiness-NML.
    \item it includes Luckiness NML as a special case, if the prior $w$ is chosen properly as a function of $\alpha$ (see Subsection C later on).
\end{itemize}

\subsection{Existence of $\alpha$-NML and choice of $w$ and $\alpha$}
The main advantage of $\alpha$-NML is that, while it is an approximation of NML that gets more and more accurate as $\alpha\to\infty$, $\alpha$-NML may exist for finite $\alpha$ even when the NML does not converge. In fact, the existence of $\alpha$-NML is determined by the convergence of the integral at the denominator of \eqref{AlphaNML}, i.e.,
\begin{equation}
\label{AlphaNMLDen}
\int_{\mathcal{X}^n} \left\{\int_{\Theta}w(\theta)\,p_{\theta}^{\alpha}(x^n)\,d\theta\right\}^{1/\alpha}\,dx^n\, .
\end{equation}
General conditions for its convergence are hard to find. However, the following sufficient conditions can be derived.
\begin{thm}
    For a given $\alpha \geq 1$, the $\alpha$-NML predictor exists if any of the following conditions are satisfied.
    \begin{enumerate}
    \item If the NML exists, so does the $\alpha$-NML, for any prior distribution $w$ on $\Theta$.
    \item If $\mathcal{P}$ is an exponential family of distributions, and if $\mathrm{supp}(w)$ is an INECCSI\footnote{Following \cite{grunwald1}, an INECCSI subset of $\Theta$ is a set $\Theta_0 \subset \Theta$ such that the interior of $\Theta_0$ is not empty, and its closure is a compact subset of the interior of $\Theta$.} subset of $\Theta$, then the $\alpha$-NML always exists.
    \item If $\Theta$ is countable, then the $\alpha$-NML exists if 
    \begin{equation}
    \label{WAlphaCond}
        \sum_{\theta\in\Theta} w(\theta)^{\frac{1}{\alpha}} < \infty.
    \end{equation}
    \item If either $\Theta$ or $\mathcal{X}$ are finite, then the $\alpha$-NML always exists.
    \end{enumerate}
\end{thm}
\begin{IEEEproof}
The proof follows from the observation that the logarithm of the denominator of $\alpha$-NML, i.e., Equation \eqref{AlphaNMLDen}, is Sibson's mutual information of order $\alpha$ (minus the multiplicative constant). The three propositions of the theorem can be proved using known properties of $I_{\alpha}(\phi, X^n)$.
\begin{enumerate}
    \item It is known that $I_{\alpha}(\phi, X^n)$ is a non-decreasing function of $\alpha$ \cite{verdu1,esposito2024sibsonsalphamutualinformationvariational}. Therefore,
    \begin{align}
        I_{\alpha}(\phi, X^n) &\leq I_{\infty}(\phi, X^n) \\
        &= \log \int_{\mathcal{X}} \sup_{\theta\in\mathrm{supp}(w)} p_{\theta}(x^n)\, dx^n \\
        &\leq \log \int_{\mathcal{X}} \sup_{\theta\in\Theta} p_{\theta}(x^n)\, dx^n.
    \end{align}
    Since the last quantity is precisely the logarithm of the denominator of the NML, then, if the NML exists, i.e., if its denominator is finite, so does $\alpha$-NML.
    \item If $\mathrm{supp}(w) \subset \Theta$ is INECCSI, then
    \begin{align}
    I_{\alpha}(\phi, X^n) &\leq I_{\infty}(\phi, X^n) \\
        &= \log \int_{\mathcal{X}} \sup_{\theta\in\mathrm{supp}(w)} p_{\theta}(x^n)\, dx^n
    \end{align}
    and the last quantity is known to be finite from \cite[Theorem 7.1]{grunwald1}.
    \item Due to the data processing inequality for Sibson's mutual information \cite{verdu1,esposito2024sibsonsalphamutualinformationvariational}, we have that $I_{\alpha}(\phi, X^n) \leq I_{\alpha}(\phi, \phi)$. Furthermore, for countable $\Theta$, \cite{verdu1} also shows that
    \begin{equation}
        I_{\alpha}(\phi, \phi) = H_{\frac{1}{\alpha}}(\phi) = \frac{\alpha}{\alpha - 1} \log \sum_{\theta} w(\theta)^{\frac{1}{\alpha}}.
    \end{equation}
    Hence, if $\sum_{\theta} w(\theta)^{\frac{1}{\alpha}}$ is finite, then $I_{\alpha}(\phi, X^n)$ is finite and the $\alpha$-NML exists.
    \item If $\Theta$ is finite, then the existence of $\alpha$-NML is guaranteed by point 3), since the sum in \eqref{WAlphaCond} contains finitely many bounded terms. If $\mathcal{X}$ is finite, then we have
    \begin{align}
        \sum_{x^n} \left\{\int_{\Theta}w(\theta)\,p_{\theta}^{\alpha}(x^n)\,d\theta\right\}^{1/\alpha} &\leq \sum_{x^n} \left\{\int_{\Theta}w(\theta)\,d\theta\right\}^{1/\alpha} \\
        &= \lvert\mathcal{X}\rvert^n
    \end{align}
    where the inequality is due to the fact that for finite $\mathcal{X}$ we have $p_{\theta}(x^n) \leq 1$ for every $\theta$ and $x^n$, and the final equality is due to the fact that $w$ is a probability distribution on $\Theta$.
\end{enumerate}
\end{IEEEproof}

Theorem 2 helps us to answer the two questions that arose in Section \ref{ANML}. In fact, the theorem highlights the fundamental role of the prior distribution $w$ on the existence of $\alpha$-NML. Even when the NML does not exist, one can carefully choose $w$ so that the $\alpha$-NML converges. In particular, condition 3) shows an interesting interplay between the prior $w$  and the parameter $\alpha$, when the parameter space $\Theta$ is countable. Note that the sum in \eqref{WAlphaCond} is a non-decreasing function of $\alpha$: when $\alpha = 1$, the sum always converges since $w$ is a probability distribution, while it diverges in the limit $\alpha\to\infty$. Since the $\alpha$-NML gets closer to the NML as $\alpha$ grows, Equation \eqref{WAlphaCond} suggests that, for a given $w$, one should choose the largest possible $\alpha$ such that $\sum_{\theta} w(\theta)^{1/\alpha}$ converges. 

In general, however, good choices of $w$ and $\alpha$ will depend on the detailed structure of parameter space $\Theta$ and the associated probability distributions. For example, for the special case $\alpha=1,$ we can connect to the classic literature on mixture estimators. Namely, when the parametric family under consideration is the class of discrete memoryless sources, one retrieves well-known estimators depending on the chosen prior $w$: when $w$ is the uniform distribution, one obtains the Laplace estimator \cite{davisson1,rissanen1}, while the Krichevsky-Trofimov estimator is obtained when $w$ is a Dirichlet distribution $D(\frac{1}{2},\dots,\frac{1}{2})$. The NML predictor is instead retrieved in the limit $\alpha \to \infty$, provided that for every $x^n\in\mathcal{X}^n$, $\sup_{\theta} p_{\theta}(x^n)$ is achieved for a $\theta$ such that $w(\theta) > 0$. This condition is achieved in particular for a prior $w$ such that $w(\theta) > 0$ for every $\theta\in\Theta$.

It is important to point out that the NML does not exist for most parametric families with an infinite parameter space $\Theta$ \cite{grunwald1}. By contrast, the $\alpha$-NML may very well exist.
In the sequel, we illustrate this fact by the aid of a number of examples.
Consider for example the case where the distribution $p_{\theta}(x^n)$ is unbounded for some $x^n \in\mathcal{X}^n$ and $\theta\in\Theta$. It is clear that the NML never exists in such a case, since $\sup_{\theta}p_{\theta}(x^n) = +\infty$ for some $x^n$. However, the $\alpha$-NML may exist nonetheless, for proper choices of the prior $w$, as in the following example. 
\begin{ex}
Consider the location family with parameter space $\Theta={\mathbb R}$ and associated distributions
\begin{equation}
p_{\theta}(x) = \begin{cases}
    -\log (x - \theta), & \theta < x \leq 1+\theta \\
    0, &\text{otherwise}.
\end{cases}
\end{equation}
While it is easy to check that this is a valid probability density function for every $\theta\in {\mathbb R},$ the key observation in this example is that $\sup_{\theta} p_{\theta}(x) = +\infty$ for every $x\in\mathbb{R}$. Therefore, the NML does not exist. However, consider the $\alpha$-NML predictor with prior $w(\theta) = \mathbbm{1}_{\{0\leq\theta\leq 1\}}$ and, e.g., $\alpha=2$. Then, the $\alpha$-NML exists. In fact, after some algebra, one can check that the integral in \eqref{AlphaNMLDen} equals
\begin{align}
\int_{\mathcal{X}^n} &\bigg\{\int_{\Theta}w(\theta)\,p_{\theta}^{\alpha}(\bar{x}^n)\,d\theta\bigg\}^{1/\alpha}\,d\bar{x}^n\notag\\
    &= \int_0^1 \Big(\sqrt{x^2\ln^2x - 2x\ln x + 2x} \notag\\
    &\hspace{5em}+ \sqrt{2 - x^2\ln^2x + 2x\ln x - 2x}\Big)\,dx \\
    &\approx 1.68.
\end{align}
Note that Luckiness NML as defined in \cite{grunwald1}, which is another alternative predictor to NML, does not exist either in this case, since for any $x\in(0,1)$, $\sup_{\theta} w(\theta)p_{\theta}(x) = +\infty$.
\end{ex}

The following is an example of a class of bounded distributions, with countable alphabet $\mathcal{X}$, for which the NML does not exist, but the $\alpha$-NML does for every $\alpha \geq 1$.
\begin{ex}
Consider the family of geometric distributions
\begin{equation}
p_{\theta}(x) = (1-\theta)^x \theta
\end{equation}
for $x \in \mathcal{X}=\mathbb{N}$. A simple calculation shows that
\begin{equation}
\sum_x\sup_{\theta} p_{\theta}(x) = \sum_x \frac{1}{x} \left(1 - \frac{1}{x+1}\right)^{x+1} = +\infty.
\end{equation}
However, consider the $\alpha$-NML with the prior $w$ as the uniform distribution on $[0,1]$ and $\alpha$ as any positive integer. Then, 
\begin{equation}
\int_0^1 w(\theta)p_{\theta}^{\alpha}(x)\, dx = \frac{1}{\alpha(x+1)+1}\cdot\frac{1}{\binom{\alpha(x+1)}{\alpha}}
\end{equation}
due to properties of the Beta function. Now, since $\binom{\alpha(x+1)}{\alpha} \sim \frac{(\alpha(x+1))^{\alpha}}{\alpha !}$ for any fixed $\alpha\geq1$ and large $x$, one has
\begin{multline}
\sum_x\left\{\int_0^1 w(\theta)p_{\theta}^{\alpha}(x)\, dx\right\}^{1/\alpha}\\
=\sum_x \left\{\frac{1}{\alpha(x+1)+1}\frac{1}{\binom{\alpha(x+1)}{\alpha}}\right\}^{1/\alpha}.
\end{multline}
This series converges if and only if the series
\begin{equation}
\sum_x \frac{1}{x^{1+\frac{1}{\alpha}}}
\end{equation}
does, which is the case for every $1\leq\alpha <\infty$. Note that with this choice of prior, Luckiness NML does not exist, either, since it is equal to the NML.
\end{ex}
Finally, the following is an example of a parametric class with uncountable $\mathcal{X}$ and $\Theta$ for which the NML does not exist, but the $\alpha$-NML does, for every $\alpha \geq 1$ and with a prior distribution $w$ such that $\mathrm{supp}(w) = \Theta$.
\begin{ex}
Consider the normal location family with variance $1$, where $\Theta = \mathbb{R}$ and $p_{\theta}(x) = \frac{1}{\sqrt{2\pi}}e^{-\frac{(x-\theta)^2}{2}}$, and consider the prior distribution $w(\theta) = \frac{1}{\sqrt{2\pi}} e^{-\frac{\theta^2}{2}}$. Then, we have
\begin{align}
\int_{\Theta} w(\theta) p_{\theta}^{\alpha}(x)\,d\theta &= \frac{1}{2\pi} \int_{-\infty}^{\infty} e^{-\frac{\theta^2}{2} - \frac{\alpha}{2}(x-\theta)^2} \\
    &= \frac{1}{2\pi} e^{-\frac{\alpha}{2(1+\alpha)}x^2} \int_{-\infty}^{\infty} e^{-\frac{1+\alpha}{2}(\theta - \frac{\alpha x}{1+\alpha})^2} \\
    &= \frac{1}{\sqrt{2\pi (1+\alpha)}} e^{-\frac{\alpha}{2(1+\alpha)}x^2} .
\end{align}
Hence, we have
\begin{align}
\int_{\mathcal{X}} \bigg\{\int_{\Theta} w(\theta)\,&p_{\theta}^{\alpha}(x)\,d\theta\bigg\}^{1/\alpha}\,dx \\
&= \frac{1}{(2\pi(1+\alpha))^{\frac{1}{2\alpha}}} \int_{-\infty}^{\infty} e^{-\frac{x^2}{2(1+\alpha)}}\, dx \\
    &= (2\pi(1+\alpha))^{\frac{1}{2}(1 - \frac{1}{\alpha})}
\end{align}
which is finite for every $1 \leq \alpha < \infty$. As expected, its value goes to infinity as $\alpha \to \infty$. In fact, for this parametric class, the NML does not exist, since $\sup_{\theta} p_{\theta}(x) = \frac{1}{\sqrt{2\pi}}$ is constant, and therefore its integral over $\mathcal{X}=\mathbb{R}$ is infinite. 
\end{ex}

\subsection{Worst-case regret of $\alpha$-NML}
As we mentioned earlier, the NML is the optimal predictor under worst-case regret \eqref{RmaxDiv}, i.e., $\alpha$-regret for $\alpha = \infty$. It is therefore interesting to study the worst-case regret of $\alpha$-NML, in order to assess how much one loses under this metric as a function of $\alpha \geq 1$. The following formula highlights the depencence of the worst-case regret for the $\alpha$-NML predictor on Sibson's $\alpha$-mutual information.

\begin{lem}
The worst-case regret of the $\alpha$-NML predictor with prior $w$ can be written as
\begin{equation}
\label{RMaxAlphaNML}
R_{\max}(\hat{p}_{\alpha}) = \frac{\alpha -1}{\alpha} I_{\alpha}(\phi, X^n) + W_{\alpha}(\mathcal{P})
\end{equation}
where $I_{\alpha}(\phi,X^n)$ is the $\alpha$-mutual information for $(\phi,X^n) \sim w(\phi)\,p_{\phi}(X^n)$, and
\begin{equation}
\label{WTerm}
W_{\alpha}(\mathcal{P}) \triangleq \sup_{x^n\in\mathcal{X}^n} \log\frac{\sup_{\theta\in\Theta} p_{\theta}(x^n)}{\left\{\int_{\Theta} w(\theta)\,p_{\theta}^{\alpha}(x^n)\,d\theta\right\}^{1/\alpha}}.
\end{equation}
\end{lem}
\begin{IEEEproof}
Starting from \eqref{WCR} and substituting the definition of $\alpha$-NML given by Equation \eqref{AlphaNML}, we have
\begin{align}
R_{\max}(\hat{p}_{\alpha}) &= \sup_{\theta\in\Theta} \sup_{x^n\in\mathcal{X}^n} \log\frac{p_{\theta}(x^n)}{\hat{p}_{\alpha}(x^n)} \\
	&= \sup_{\theta\in\Theta} \sup_{x^n\in\mathcal{X}^n} \log\frac{p_{\theta}(x^n)}{\frac{\left\{\int_{\Theta}w(\theta)\,p_{\theta}^{\alpha}(x^n)\,d\theta\right\}^{1/\alpha}}{\int_{\mathcal{X}^n} \left\{\int_{\Theta}w(\theta)\,p_{\theta}^{\alpha}(\bar{x}^n)\,d\theta\right\}^{1/\alpha}\,d\bar{x}^n}} \\
	&= \log \int_{\mathcal{X}^n} \left\{\int_{\Theta}w(\theta)\,p_{\theta}^{\alpha}(x^n)\,d\theta\right\}^{1/\alpha}dx^n \notag\\
    &\hspace{4em}+ \max_{x^n} \log\frac{\sup_{\theta} p_{\theta}(x^n)}{\left\{\int_{\Theta}w(\theta)\,p_{\theta}^{\alpha}(x^n)\,d\theta\right\}^{1/\alpha}} \\
	&= \frac{\alpha -1}{\alpha} I_{\alpha}(\phi, X^n) + W_{\alpha}(\mathcal{P})
\end{align}
where in the last step we used the definitions of $I_{\alpha}(\phi, X^n)$ and $W_{\alpha}(\mathcal{P})$ in Equations \eqref{IAlphaDef} and \eqref{WTerm} respectively.
\end{IEEEproof}

Since in the limit $\alpha \to\infty$ the $\alpha$-NML predictor becomes equal to the NML, it follows that $R_{\max}(\hat{p}_{\alpha})$ tends to the optimal worst-case regret $I_{\infty}(\phi, X^n)$ when $\alpha$ goes to infinity. However, in general, it is not clear from \eqref{RMaxAlphaNML} what is the behavior of $R_{\max}(\hat{p}_{\alpha})$ as a function of $\alpha$, i.e., it is not clear if the regret is monotonically decreasing with $\alpha$ or not, and this might depend on the actual class of distributions that is considered. In fact, the first term in \eqref{RMaxAlphaNML} is increasing with $\alpha$, due to known properties of Sibson's $\alpha$-mutual information \cite{verdu1,esposito2024sibsonsalphamutualinformationvariational}. However, the overall behavior of the regret is certainly not increasing with $\alpha$, since it reaches its minimum when $\alpha \to \infty$. This proves the critical role of the second term $W_{\alpha}(\mathcal{P})$ in the overall behavior of the worst-case regret. While this term is in general of difficult analysis, in Section \ref{ANMLDMS} we show that it can be written in a simple form for the class of discrete memoryless sources, for which a complete asymptotic analysis of the worst-case regret can be carried out.

\subsection{Connection between $\alpha$-NML and Luckiness NML}
The previous sections already made clear that the choice of the prior distribution $w$ in \eqref{AlphaNML} in defining the $\alpha$-NML distribution is of fundamental importance. For example, the choice of a Dirichlet prior in \eqref{AlphaNMLDMC} is what makes $\alpha$-NML almost optimal for the case of discrete memoryless sources, and the choice of the correct prior is necessary for the optimality of $\alpha$-NML under the $\alpha$-regret \eqref{LambdaRegret}. When we discussed in Section \ref{ANML} that the $\alpha$-NML interpolates the mixture predictors \eqref{MixtureDef} and the NML \eqref{NML}, we assumed $w$ to be fixed and independent of $\alpha$. It turns out that if one chooses $w$ carefully as a function of $\alpha$, the $\alpha$-NML can also approximate other predictors related to the NML, in particular the Luckiness NML predictor defined in \eqref{LNML}.
In fact, let $\pi$ be the luckiness function used in the definition of Luckiness NML, and consider as a regret measure the expectation over the parameters in $\Theta$ according to $\pi$, and then the expectation over sequences distributed according to $p_{\theta}$. The result is what we may call \emph{average luckiness regret}, which is formally defined as
\begin{equation}
\label{AvLRegret}
R_{\rm av}(\pi,\hat{p}) = \mathbb{E}_{\theta \sim\pi}\left[\mathbb{E}_{X^n\sim p_{\theta}}\left[\log\frac{p_{\theta}(X^n)}{\hat{p}(X^n)}\right]\right].
\end{equation}
It is easy to prove that the predictor minimizing this regret is a mixture predictor whose weighting function is $\pi$, i.e., 
\begin{equation}
 \label{LMixture}
 \hat{p}(x^n) = \int_{\Theta} \pi(\theta) p_{\theta}(x^n)d\theta\,.
\end{equation}
A possible interpolation between the Luckiness NML defined in \eqref{LNML} and the mixture predictor in \eqref{LMixture} is again given by $\alpha$-NML of Equation \eqref{AlphaNML}, if one chooses the proper prior distribution $w$. In fact, for any given $\alpha \geq 1$, one can take the tilted prior distribution
\begin{equation}
\label{LTiltedPrior}
w(\theta) = \frac{\pi(\theta)^{\alpha}}{\int_{\Theta} \pi(\theta)^{\alpha} d\theta}
\end{equation}
provided that the integral in the denominator converges. With such a choice of prior, the $\alpha$-NML becomes
\begin{align}
\hat{p}_{\alpha}(x^n) &= \frac{\left\{\int_{\Theta}w(\theta)\,p_{\theta}^{\alpha}(x^n)\,d\theta\right\}^{1/\alpha}}{\sum_{x^n} \left\{\int_{\Theta}w(\theta)\,p_{\theta}^{\alpha}(x^n)\,d\theta\right\}^{1/\alpha}} \\
	&= \frac{\left\{\int_{\Theta}(\pi(\theta)\,p_{\theta}(x^n))^{\alpha}\,d\theta\right\}^{1/\alpha}}{\sum_{x^n} \left\{\int_{\Theta}(\pi(\theta)\,p_{\theta}(x^n))^{\alpha}\,d\theta\right\}^{1/\alpha}}. \label{LAlphaNML}
\end{align}
For convenience, we can call this predictor \emph{Luckiness $\alpha$-NML}. However, it is important to note that this predictor is not something different from the already defined $\alpha$-NML. In fact, it is simply a particular instance of that same predictor, where one chooses a particular prior distribution $w$ -- in this case, it is the one in Equation \eqref{LTiltedPrior}. In this sense, the $\alpha$-NML is able to link the standard NML and the Luckiness NML under the same, more general, object.
By taking $\alpha=1$, one retrieves the mixture predictor in \eqref{LMixture}, while in the limit $\alpha \to \infty$, one gets the luckiness NML that is defined in \eqref{LNML}. 

Similarly to the case of the $\alpha$-regret of Equation \eqref{LambdaRegret}, there also exists an interpolation between the worst-case luckiness regret in \eqref{WCLRegret} and the average luckiness regret in \eqref{AvLRegret} for which the Luckiness $\alpha$-NML is the optimal predictor. In fact, consider the luckiness regret defined for any $\alpha\geq 1$ by
\begin{equation}
\label{LRLambda}
R_{\alpha}(\pi,\hat{p}) = \frac{1}{\alpha - 1}\log\mathbb{E}_{\theta\sim\pi_{\alpha}}\left[\mathbb{E}_{X^n\sim p_{\theta}}\left[\left(\frac{p_{\theta}(X^n)}{\hat{p}(X^n)}\right)^{\alpha-1}\right]\right]
\end{equation}
where
\begin{equation}
\pi_{\alpha}(\theta) = \frac{\pi(\theta)^{\alpha}}{\int_{\Theta}\pi(\bar{\theta})^{\alpha}d\bar{\theta}}.
\end{equation}
Notice that the exponent of $\alpha$ inside the expectation is the same as in \eqref{LambdaRegret}, as well as the normalization factor $\frac{1}{\alpha -1}\log$ in front. However, while the interpolation of $\alpha$-regret discussed in Section \ref{ARegret} acts only on how the sequences $x^n$ are considered, in the luckiness case, instead, the interpolation also occurs on how the parameters in $\Theta$ are weighted. In fact, in the worst-case luckiness regret \eqref{WCLRegret} there is a maximization over $\theta$, while in the average luckiness regret \eqref{AvLRegret} there is an expectation according to $\pi$. The way this interpolation is handled in \eqref{LRLambda} is through an expectation over a tilted version of the luckiness function $\pi$, which equals $\pi$ when $\alpha = 1$, and it assigns probability one to the maximal $\theta$ when $\alpha\to\infty$. The optimal predictor for the luckiness $\alpha$-regret is the Luckiness $\alpha$-NML.
\begin{thm}
For any given $\alpha\geq 1$ and any given luckiness function $\pi$ over $\Theta$, the Luckiness $\alpha$-NML defined in \eqref{LAlphaNML} with prior distribution $w$ taken as in \eqref{LTiltedPrior}, 
minimizes $R_{\alpha}(\pi,\hat{p})$ over all probability distributions on $\mathcal{X}^n$, under the assumption that the prior distribution converges, i.e., if
\begin{equation}
\int_{\Theta} \pi(\theta)^{\alpha} d\theta < \infty.
\end{equation}
\end{thm}
\begin{IEEEproof}
See Appendix \ref{Proof3}.
\end{IEEEproof}

Alternatively, one could define a different average luckiness regret measure such as
\begin{equation}
\label{AvLRegret2}
\tilde{R}_{\rm av}(\pi,\hat{p}) = \max_{\theta\in\Theta}\mathbb{E}_{X^n\sim p_{\theta}}\left[\log\frac{\pi(\theta) p_{\theta}(X^n)}{\hat{p}(X^n)}\right].
\end{equation}
Then, a simple interpolation between this regret and \eqref{WCLRegret} is
\begin{equation}
\label{LRLambda2}
R_{\alpha}(\pi,\hat{p}) = \sup_{\theta\in\Theta} \frac{1}{\alpha -1} \log \mathbb{E}_{X^n\sim p_{\theta}} \left[\left(\frac{\pi(\theta)p_{\theta}(X^n)}{\hat{p}(X^n)}\right)^{\alpha-1}\right]
\end{equation}
which is strongly related to the $\alpha$-regret in Equation \eqref{LambdaRegret}. In fact, one can prove a result similar to Theorem 3 for this regret measure. In this context, it can be used to show that there exists a distribution $w^*(\theta)$ on $\Theta$ such that, the $\alpha$-NML defined in \eqref{AlphaNML}, with prior equal to
\begin{equation}
w(\theta) = \frac{w^*(\theta)\pi^{\alpha}(\theta)}{\int_{\Theta}w^*(\bar{\theta})\pi^{\alpha}(\bar{\theta})d\bar{\theta}}
\end{equation}
is the predictor that minimizes \eqref{LRLambda2}.


\section{$\alpha$-NML for DMS}
\label{ANMLDMS}
We now focus on the important class of discrete memoryless sources taking values in a finite but arbitrary alphabet\footnote{In part of the literature this class also goes under the name of \emph{constant experts} --- see, e.g., \cite{cesa1}.}. This class has been the focus of a large part of the literature on universal prediction and compression. The main reasons for this are that this class is the simplest non-trivial example for which one can get a sense of how a predictor behaves, and at the same time prove rigorously some results in terms of performance of a predictor compared to the optimal.

\subsection{The Krichevsky-Trofimov estimator}
The most important result on universal prediction for this class of distributions is possibly the Krichevsky-Trofimov estimator. Let the source alphabet be $\mathcal{X} = \{1,2,\dots,m\}$. Let also 
\begin{equation}
\Theta = \left\{\bm{\theta} = (\theta_1,\theta_2,\dots,\theta_m) : \sum_{i=1}^m \theta_i = 1 \text{ and } \theta_i \geq 0 \ \forall i\right\}
\end{equation}
be the parameter set. For each parameter $\bm{\theta}$ and sequence $x^n=(x_1,x_2,\dots,x_n) \in \mathcal{X}^n$, the source indexed by $\bm{\theta}$ generates the sequence $x^n$ with probability
\begin{equation}
p_{\bm{\theta}}(x^n) = \prod_{i=1}^m \theta_i^{n_i}\,,
\end{equation}
where 
\begin{equation}
n_i = \left\lvert\left\{ 1\leq j\leq n : x_j = i\right\}\right\rvert.
\end{equation}

For the class of discrete memoryless sources described above, the Krichevsky-Trofimov predictor is a simple mixture estimator,
\begin{equation}
\label{KTDefInt}
\hat{p}_{\rm KT}(x^n) \triangleq \int_{\Theta} w_{\rm D}(\bm{\theta})\,p_{\bm{\theta}}(x^n)\,d\bm{\theta}
\end{equation}
where the prior distribution on the parameter space is $w_{\rm D} \sim D(\frac{1}{2},\dots,\frac{1}{2})$, i.e., the Dirichlet distribution with parameters equal to $\frac{1}{2}$,
\begin{equation}
\label{WD}
w_{\rm D}(\bm{\theta}) = \frac{\Gamma\left(\frac{m}{2}\right)}{\pi^{m/2}}\prod_{i=1}^m \frac{1}{\sqrt{\theta_i}}\,.
\end{equation}
This estimator has arguably three major advantages.
\begin{enumerate}
\item Its probability estimates $\hat{p}_{\rm KT}(x^n)$ can be computed easily in closed form. In fact, substituting the definitions of $w_{\rm D}$ and of $p_{\bm{\theta}}$ into \eqref{KTDefInt} and using properties of the Gamma function $\Gamma(t)$, one is able to derive the simple formula
\begin{equation}
\hat{p}_{\rm KT}(x^n) = \frac{\Gamma\left(\frac{m}{2}\right)}{\pi^{m/2}} \frac{\prod_{i=1}^m \Gamma\left(n_i + \frac{1}{2}\right)}{\Gamma\left(n + \frac{m}{2}\right)}\,.
\end{equation}
\item It is asymptotically optimal in $n$ up to a constant term, in terms of both worst-case regret $R_{\max}$ and average regret $R_{\rm av}$.
\item Simple formulae exist for the computation of the conditional probability of a new symbol given the previous ones.
\end{enumerate}


\subsection{The $\alpha$-NML for DMS}

We now compare the Krichevsky-Trofimov estimator with the $\alpha$-NML for the class of discrete memoryless sources. We will use as a prior distribution $w$ for the $\alpha$-NML the same Dirichlet $D(\frac{1}{2},\dots,\frac{1}{2})$ of the KT estimator. It is important to notice that other prior distributions could also be considered. In this work, we focus on the Dirichlet prior distribution mainly for two reasons: (1) it makes easier to compare our predictor to the Krichevsky-Trofimov, since with such a choice, the $\alpha$-NML is precisely the KT predictor when $\alpha=1$; (2) it is easier to handle mathematically and to get closed-form formulas for the estimated probabilities. Furthermore, the Dirichlet distribution $D(\frac{1}{2},\dots,\frac{1}{2})$ is the so-called \emph{Jeffreys' prior distribution} \cite{jeffreys1} for the class of discrete memoryless sources. It is known  that a mixture predictor with prior distribution equal to Jeffrey's prior has an asymptotically optimal regret, for exponential families of distributions and for most sequences $x^n$ in $\mathcal{X}^n$ (see, e.g., \cite[Section 8.1]{grunwald1} and references therein, for a more precise account of these results).
Nevertheless, it is likely that other prior distributions would improve the performance of the predictor, at the cost of additional complexity of implementation.

In the case of the Dirichlet distribution $w_D$ as in \eqref{WD}, the $\alpha$-NML predictor takes the form
\begin{equation}
\label{AlphaNMLDMC}
\hat{p}_{\alpha}(x^n) = \frac{1}{Z_n(\alpha)}\left\{\int_{\Theta} \prod_{i=1}^m \theta_i^{\alpha n_i - \frac{1}{2}}\,d\bm{\theta}\right\}^{1/\alpha}\,,
\end{equation}
where the normalization constant $Z_n(\alpha)$ is equal to
\begin{equation}
Z_n(\alpha) \triangleq \sum_{x^n} \left\{\int_{\Theta} \prod_{i=1}^m \theta_i^{\alpha n_i - \frac{1}{2}}\,d\bm{\theta}\right\}^{1/\alpha}\,.
\end{equation}
The integral on the right is known in the literature as the multivariate Beta function, and it has the closed-form expression
\begin{equation}
\int_{\Theta} \prod_{i=1}^m \theta_i^{\alpha n_i - \frac{1}{2}}\,d\bm{\theta} = \frac{\prod_{i=1}^m \Gamma\left(\alpha n_i + \frac{1}{2}\right)}{\Gamma\left(\alpha n + \frac{m}{2}\right)}\,,
\end{equation}
so that the probability estimates given by the $\alpha$-NML predictor can be written as
\begin{equation}
\label{PAlphaDef}
\hat{p}_{\alpha}(x^n) = \frac{1}{Z_n(\alpha)} \left\{\frac{\prod_{i=1}^m \Gamma\left(\alpha n_i + \frac{1}{2}\right)}{\Gamma\left(\alpha n + \frac{m}{2}\right)}\right\}^{1/\alpha}
\end{equation}
where
\begin{equation}
\label{ZAlphaDef}
Z_n(\alpha) = \sum_{x^n} \left\{\frac{\prod_{i=1}^m \Gamma\left(\alpha n_i + \frac{1}{2}\right)}{\Gamma\left(\alpha n + \frac{m}{2}\right)}\right\}^{1/\alpha}\,.
\end{equation}

We now want to briefly discuss the computational complexity of $\alpha$-NML. Notice that in principle the sum that appears in $Z_n(\alpha)$ contains an exponential number of terms in $n$, which may be of concern from a computational point of view. However, it can be seen that the actual terms in the sum only depend on the number of symbols $\bm{n} = (n_1,n_2,\dots,n_m)$. Therefore, one can group equal terms together to get
\begin{equation}
Z_n(\alpha) = \sum_{\bm{n}} \binom{n}{n_1,\dots,n_m}\left\{\frac{\prod_{i=1}^m \Gamma\left(\alpha n_i + \frac{1}{2}\right)}{\Gamma\left(\alpha n + \frac{m}{2}\right)}\right\}^{1/\alpha}\,.
\end{equation}
Written in this way, the sum contains only a polynomial number of terms, since the number of different vectors $\bm{n}$ is upper-bounded by $(n+1)^{m-1}$. In particular, when the alphabet is binary --- i.e., when $m=2$, --- the number of terms is linear in $n$. Furthermore, the computation of the multinomial coefficients is also not a problem, since they can be computed recursively from the previous ones with a constant number of operations.

Finally, the Gamma terms in \eqref{PAlphaDef} and \eqref{ZAlphaDef} can also be computed efficiently, when $\alpha \geq 1$ is restricted to be an integer. In such a case, one can use the recurrence formula for the Gamma function
\begin{equation}
\label{Recurrence}
\Gamma(z + 1) = z\Gamma(z)
\end{equation}
to compute each of the Gamma terms in the two formulae, e.g.,
\begin{equation}
\Gamma\left(\alpha n_i + \frac{1}{2}\right) = \left(\alpha n_i - \frac{1}{2}\right)\left(\alpha n_i - \frac{3}{2}\right)\cdots\frac{3}{2}\cdot\frac{1}{2}\cdot\sqrt{\pi}\,,
\end{equation}
where we used the well-known fact that $\Gamma\left(\frac{1}{2}\right) = \sqrt{\pi}$. Similar computations can be used to calculate the denominator term $\Gamma\left(\alpha n + \frac{m}{2}\right)$. As one can see, the number of operations required for each term of the sum in \eqref{ZAlphaDef} is linear in $\alpha n$. Therefore, for any positive integer $\alpha$, the number of operations required to compute $Z_n(\alpha)$ and $\hat{p}_{\alpha}(x^n)$ is polynomial in $n$ and linear in $\alpha$. As we will see later on, a small value of $\alpha$ is already enough to improve significantly the worst-case regret of the $\alpha$-NML predictor, and to get close to the optimal regret achieved by the NML.

When $\alpha$ is a positive integer, one can also derive simple formulae for the conditional probability of the next symbol when a sequence of length $n-1$ is already given. Consider the setting where a fixed sequence $x^{n-1} \in \mathcal{X}^{n-1}$ has been revealed, and we want to estimate the conditional probability of symbol $k \in \mathcal{X}$ given $x^{n-1}$, where $\mathcal{X} = \{1,2,\dots,m\}$. As an intermediate step, let us compute the ratio $\hat{p}_{\alpha}(x^{n-1}, k)/\hat{p}_{\alpha}(x^{n-1})$.
\begin{align}
\frac{\hat{p}_{\alpha}(x^{n-1},k)}{\hat{p}_{\alpha}(x^{n-1})} &= \frac{\frac{1}{Z_{n}(\alpha)}\left\{\frac{\Gamma(\alpha (n_k + 1)+\frac{1}{2})\prod_{i\neq k}\Gamma(\alpha n_i + \frac{1}{2})}{\Gamma(\alpha n + \frac{m}{2})}\right\}^{\frac{1}{\alpha}}}{\frac{1}{Z_{n-1}(\alpha)}\left\{\frac{\Gamma(\alpha n_k + \frac{1}{2})\prod_{i\neq k}\Gamma(\alpha n_i + \frac{1}{2})}{\Gamma(\alpha (n-1) + \frac{m}{2})}\right\}^{\frac{1}{\alpha}}} \\
	&= \frac{Z_{n-1}(\alpha)}{Z_{n}(\alpha)} \left\{\prod_{j=0}^{\alpha-1}\frac{\alpha n_k + \frac{1}{2} + j}{\alpha n -\alpha+ \frac{m}{2} + j}\right\}^{\frac{1}{\alpha}},
\end{align}
where in the last step we used \eqref{Recurrence} recursively. Finally, we can obtain the conditional probability of $k$ given $x^{n-1}$ as
\begin{align}
\hat{p}_{\alpha}(k | x^{n-1}) &\triangleq \frac{\hat{p}_{\alpha}(x^{n-1}, k)}{\sum_{i=1}^m \hat{p}_{\alpha}(x^{n-1}, i)} \\
	&= \frac{\frac{\hat{p}_{\alpha}(x^{n-1}, k)}{\hat{p}_{\alpha}(x^{n-1})}}{\sum_{i=1}^m \frac{\hat{p}_{\alpha}(x^{n-1}, i)}{\hat{p}_{\alpha}(x^{n-1})}} \\
	&= \frac{\prod_{j=0}^{\alpha-1}(\alpha n_k + \frac{1}{2} + j)^{1/\alpha}}{\sum_{i=1}^m \prod_{j=0}^{\alpha-1}(\alpha n_i + \frac{1}{2} + j)^{1/\alpha}} \label{CondProbFinal}
\end{align}
for any $k \in \mathcal{X}$. As one can see from \eqref{CondProbFinal}, the computational complexity of each of these probabilities is linear in $\alpha$ and $m$ and does not depend on $n$. For $\alpha =1$, one obtains the known formula for the conditional probabilities of the Krichevsky-Trofimov estimator
\begin{equation}
\hat{p}_{\rm KT}(k|x^{n-1}) = \frac{n_k + \frac{1}{2}}{n + \frac{m}{2}-1}.
\end{equation}
while, e.g., for $\alpha=2$, one gets the formula mentioned in the Introduction in Equation \eqref{Alpha2Cond}.

\subsection{Worst-case regret of $\alpha$-NML for DMS}

\begin{figure}[!t]
\centering
\includegraphics[width=0.4\textwidth]{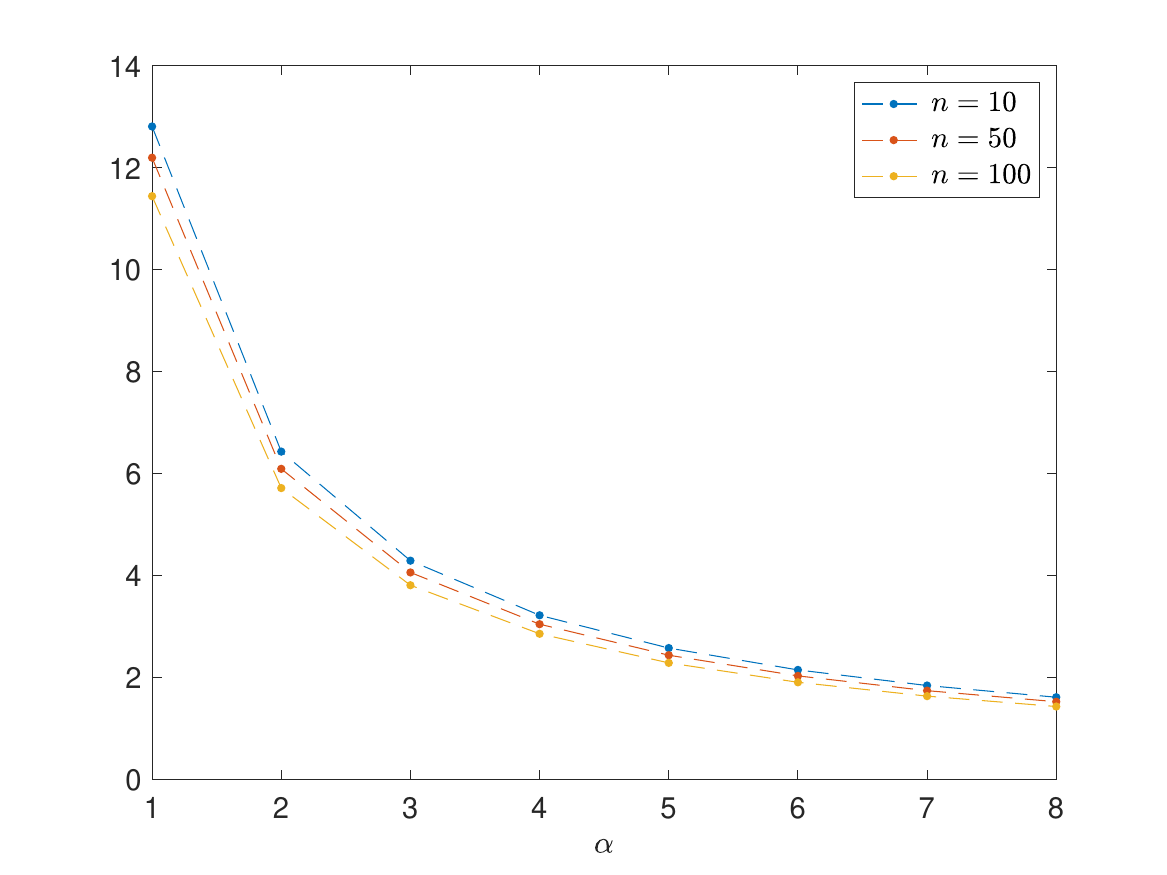}
\caption{Percentage of increase of $R_{\max}(\hat{p}_{\alpha})$ with respect to the optimal value $R_{\max}(\hat{p}_{\rm NML})$, as a function of $\alpha$, for binary sequences of length $n=10,50,100$ and integer values of $\alpha$. The value at $\alpha = 1$ corresponds to the regret of the Krichevsky-Trofimov estimator.}
\end{figure}

We now want to discuss the performance of $\alpha$-NML in terms of worst-case regret, with the primary objective of analyzing how much the regret of $\alpha$-NML improves upon that of the Krichevsky-Trofimov estimator, and how it compares to the optimal NML. In order to do this, we start by finding the asymptotical value of the worst-case regret for $\alpha$-NML, starting from formula \eqref{RMaxAlphaNML}. This formula has two major advantages in the discrete memoryless case. First, the asymptotics of the $\alpha$-mutual information term, which would be in general hard to study, can actually be computed using known results in the literature, once one recognizes the optimality of the Dirichlet prior. Second, the maximization over sequences in $\mathcal{X}^n$ in the $W_{\alpha}(\mathcal{P})$ term, that would be complicated to evaluate in general, can be resolved explicitly for this particular class of distributions.

\begin{thm}
For the class of discrete memoryless sources, the $W_{\alpha}(\mathcal{P})$ term defined in \eqref{WTerm} is equal to
\begin{equation}
\label{WTermDMS}
W_{\alpha}(\mathcal{P}) = \frac{1}{\alpha}\log\frac{\Gamma(\alpha n + \frac{m}{2})}{\Gamma(\alpha n + \frac{1}{2})} + \frac{1}{2\alpha}\log \pi - \frac{1}{\alpha}\Gamma\left(\frac{m}{2}\right)\,.
\end{equation}
\end{thm}
\begin{IEEEproof}
See Appendix \ref{Proof5}.
\end{IEEEproof}

With the help of this result, we can prove the asymptotics of the worst-case regret for the $\alpha$-NML estimator.

\begin{thm}
\label{RAlphaNMLAsymptotics}
For the class of discrete memoryless sources, the worst-case regret of the $\alpha$-NML predictor is equal to
\begin{multline}
R_{\max}(\hat{p}_{\alpha}) = \frac{m-1}{2}\log\frac{n}{2} + \frac{1}{2}\log\pi \\
-\log\Gamma\left(\frac{m}{2}\right) + \frac{m-1}{2\alpha}\log 2+ o(1) \label{RMaxAsymptotics}
\end{multline}
where $o(1) \to 0$ as $n\to\infty$.
\end{thm}
\begin{IEEEproof}
See Appendix \ref{Proof6}.
\end{IEEEproof}

From \eqref{RMaxAsymptotics} it can be seen that the asymptotic behavior of the worst-case regret of $\alpha$-NML has the same dependence on $n$ for every $\alpha \geq 1$, while the terms that do not depend on $n$ strictly decrease as $\alpha$ increases. Therefore, the $\alpha$-NML has an asymptotic advantage with respect to the Krichevsky-Trofimov estimator only in the constant term. However, for finite length, computer evaluation of the worst-case regret show that the advantage of $\alpha$-NML over the KT estimator is larger. For example, Figure 1 shows some of these results for binary alphabet. Since asymptotically the difference of the regret of the $\alpha$-NML (and in particular the Krichevsky-Trofimov estimator) and that of the NML is a constant, one expects the percentage of increase of the regret to tend to zero as $n$ goes to infinity, for every $\alpha$. However, as one can see from Figure 1, this decrease appears to be very slow, an additional indication that the (almost) optimality of the Krichevsky-Trofimov estimator in terms of worst-case regret is only asymptotical, while for finite-length sequences the difference is actually substantial. However, precise analysis of finite-length regret remains difficult.

\section{Conclusion}
In this paper, we introduced a new class of general predictors dependent on a real parameter $\alpha\geq 1$, which is shown to interpolate to mixture predictors and the NML. The idea for this class of predictors comes from the connection between the worst-case regret achieved by the NML predictor, and Sibson's $\alpha$-mutual information. We proved the optimality of $\alpha$-NML under $\alpha$-regret, a general regret measure linked to R\'{e}nyi divergence, that interpolates between the classical average and worst-case regrets. Also, we discussed examples that prove the broad applicability of $\alpha$-NML also for families of distributions for which the NML does not exist, and we compared $\alpha$-NML to other alternatives to NML such as Luckiness NML. Furthermore, we showed that for the popular family of discrete memoryless sources, one is able to derive some simple formulas to compute the probabilities estimated by the new class of predictors, when the parameter $\alpha$ is a positive integer. For this class of distributions, we also derive an asymptotic expression for the worst-case regret of $\alpha$-NML, which interpolates between those of the Krichevsky-Trofimov estimator and the NML.

\appendices

\section{Proof of Theorem 3}
\label{Proof3}
Notice that one can rewrite \eqref{LRLambda} as
\begin{equation}
R_{\alpha}(\pi,\hat{p}) = D_{\alpha}(\pi_{\alpha} \times p_{\theta} \| \pi_{\alpha} \times \hat{p}).
\end{equation}
Thanks to \cite[Equation (32)]{verdu1}, it follows that the minimum regret over all predictors $\hat{p}$ satisfies
\begin{equation}
\label{MinLRLambda}
\min_{\hat{p}} R_{\alpha}(\pi,\hat{p}) = \min_{\hat{p}} D_{\alpha}(\pi_{\alpha} \times p_{\theta} \| \pi_{\alpha} \times \hat{p}) = I_{\alpha}(\pi_{\alpha}, p_{\theta}).
\end{equation}
Furthermore, one can check by substituting the definition of luckiness $\alpha$-NML \eqref{LAlphaNML} with prior $\pi_{\alpha}$ into \eqref{LRLambda}, that the regret of the luckiness $\alpha$-NML is equal to
\begin{equation}
\label{LRLambdaNML}
R_{\alpha}(\pi,\hat{p}_{\alpha}) = I_{\alpha}(\pi_{\alpha}, p_{\theta}).
\end{equation}
From \eqref{MinLRLambda} and \eqref{LRLambdaNML} it follows that the Luckiness $\alpha$-NML minimizes the regret.

\section{Proof of Theorem 5}
\label{Proof5}
For the discrete memoryless sources case, one can rewrite \eqref{WTerm} as
\begin{align}
W_{\alpha}(\mathcal{P}) &= \max_{\bm{n}} \log\frac{\max_{\bm{\theta}} \prod_{i=1}^m \theta_i^{n_i}}{\left\{\frac{\Gamma(\frac{m}{2})}{\pi^{m/2}}\frac{\prod_{i=1}^m\Gamma(\alpha n_i + \frac{1}{2})}{\Gamma(\alpha n + \frac{m}{2})}\right\}^{1/\alpha}} \\
	&= \max_{\bm{n}} \log\frac{\prod_{i=1}^m (\frac{n_i}{n})^{n_i}}{\left\{\frac{\Gamma(\frac{m}{2})}{\pi^{m/2}}\frac{\prod_{i=1}^m\Gamma(\alpha n_i + \frac{1}{2})}{\Gamma(\alpha n + \frac{m}{2})}\right\}^{1/\alpha}} \\
	&= \frac{1}{\alpha}\log\frac{\pi^{m/2}\,\Gamma(\alpha n + \frac{m}{2})}{\Gamma(\frac{m}{2})}-n\log n \notag\\
		&\hspace{2em}+ \max_{\bm{n}} \sum_{i=1}^m \left(n_i\log n_i - \frac{1}{\alpha}\log\Gamma\left(\alpha n_i + \frac{1}{2}\right)\right)
\end{align}
where the maximization is over vectors $\bm{n} = (n_1,n_2,\dots,n_m)$ with integer entries such that $\sum_{i=1}^m n_i = n$ and $n_i \geq 0$ for every $i$. Notice that to prove the theorem, it suffices to show that the quantity
\begin{equation}
\label{LogGammaQuantity}
\sum_{i=1}^m \left(n_i\log n_i - \frac{1}{\alpha}\log\Gamma\left(\alpha n_i + \frac{1}{2}\right)\right)
\end{equation}
is maximized for $n_m = n$ and $n_i = 0$ for every $i\neq m$, for every $n \geq 1$ and $m \geq 2$. We prove this by induction on $m$. For $m=2$, let $t = \frac{n_1}{n}$, $0\leq t\leq 1$. Then, we wish to prove that the function
\begin{multline}
f(t) = nt\log (nt) - \frac{1}{\alpha}\log \Gamma\left(\alpha n t + \frac{1}{2}\right) \\
	+ n(1-t)\log (n(1-t)) - \frac{1}{\alpha}\log \Gamma\left(\alpha n (1-t) + \frac{1}{2}\right)
\end{multline}
is maximized at $t=1$ for $0\leq t\leq 1$. Notice that $f(t)$ is symmetrical around $t = \frac{1}{2}$. Hence, it suffices to prove that $f(t)$ is convex for $0\leq t\leq 1$, and to prove this it is enough to show that
\begin{equation}
g(t) = nt\log (nt) - \frac{1}{\alpha}\log \Gamma\left(\alpha n t + \frac{1}{2}\right)
\end{equation}
is convex for $0\leq t\leq 1$. Notice that
\begin{align}
g'(t) &= n\log(nt) + n - n\,\psi\left(\alpha nt + \frac{1}{2}\right) \\
	&= n - n\log\alpha + n\log(\alpha nt) - n\,\psi\left(\alpha nt + \frac{1}{2}\right) \\
	&= n - n\log\alpha + n\, h(\alpha nt)
\end{align}
where $\psi(x) = \frac{d}{dx} \log\Gamma(x)$ is the digamma function, and
\begin{equation}
h(x) = \log(x) - \psi\left(x + \frac{1}{2}\right)\,.
\end{equation}
By \cite[Theorem 4.2]{alzer1}, it follows that $h'(x) \geq 0$ for every $x\geq 0$. Therefore, one has
\begin{equation}
g''(t) = \alpha n^2 h'(\alpha nt) \geq 0
\end{equation}
for every $0\leq t\leq 1$, i.e., $g(t)$ is convex. Hence, $f(t)$ is maximized at $t=1$, and the case $m=2$ is proved. Assume now that the case $m=k$ is true, i.e., that \eqref{LogGammaQuantity} is maximized for $n_k = n$ and $n_i=0$ for $i\neq k$, for every $n\geq 1$. Consider the case $m = k+1$. For every $\bm{n} = (n_1,n_2,\dots,n_{k+1})$, one has
\begin{align}
\sum_{i=1}^m \bigg(&n_i\log n_i - \frac{1}{\alpha}\log\Gamma\left(\alpha n_i + \frac{1}{2}\right)\bigg) \label{LogGammaFirst}\\
	&= \sum_{i=1}^k \left(n_i\log n_i - \frac{1}{\alpha}\log\Gamma\left(\alpha n_i + \frac{1}{2}\right)\right) \notag\\
	&\hspace{5em}+ n_{k+1}\log n_{k+1} - \frac{1}{\alpha}\log\Gamma\bigg(\alpha n_{k+1} + \frac{1}{2}\bigg) \\
	&\leq -\sum_{i=1}^{k-1} \frac{1}{\alpha} \log\Gamma\bigg(\frac{1}{2}\bigg) + (n-n_{k+1})\log(n-n_{k+1}) \notag\\
	&\hspace{2em}- \frac{1}{\alpha}\log\Gamma\bigg(\alpha(n-n_{k+1})+\frac{1}{2}\bigg) + n_{k+1}\log n_{k+1} \notag\\
	&\hspace{12em}- \frac{1}{\alpha}\log\Gamma\bigg(\alpha n_{k+1} + \frac{1}{2}\bigg) \\
	&\leq -\sum_{i=1}^k \frac{1}{\alpha} \log\Gamma\bigg(\frac{1}{2}\bigg) + n\log n -\frac{1}{\alpha}\log\Gamma\bigg(\alpha n + \frac{1}{2}\bigg) \label{LogGammaLast}
\end{align}
where the first inequality follows from the case $m=k$, and the second inequality follows from the case $m=2$. Thus, \eqref{LogGammaLast} shows that \eqref{LogGammaFirst} is maximized for $n_{k+1} = n$, as desired. Hence, the case $m=k+1$ is proved, and the theorem follows.

\section{Proof of Theorem 6}
\label{Proof6}
We start from Equation \eqref{RMaxAlphaNML}. The asymptotics of the $\alpha$-mutual information term indirectly follows from the proof of Theorem 2 in \cite{yagli1}. In fact, the theorem states that
\begin{multline}
\label{SupIAlpha}
\sup_{w\in\mathcal{P}(\Theta)} I_{\alpha}(\phi, X^n) = \frac{m-1}{2}\log \frac{n}{2} + \frac{1}{2}\log\pi \\
-\log\Gamma\left(\frac{m}{2}\right) -\frac{m-1}{2(\alpha - 1)}\log \alpha + o(1)\,,
\end{multline}
from which it follows that
\begin{multline}
\label{IAlphaUpper}
I_{\alpha}(\phi,X^n) \leq \frac{m-1}{2}\log \frac{n}{2} + \frac{1}{2}\log\pi \\-\log\Gamma\left(\frac{m}{2}\right) -\frac{m-1}{2(\alpha - 1)}\log \alpha + o(1)
\end{multline}
for $(\phi,X^n) \sim w(\phi)\,p_{\theta}(X^n)$ and $w$ taken as the Dirichlet distribution $\mathrm{Dir}(1/2,\dots,1/2)$. However, again in \cite{yagli1}, from Equation (80) onwards they also prove that
\begin{multline}
\label{IAlphaLower}
I_{\alpha}(\phi,X^n) \geq \frac{m-1}{2}\log \frac{n}{2} + \frac{1}{2}\log\pi \\
-\log\Gamma\left(\frac{m}{2}\right) -\frac{m-1}{2(\alpha - 1)}\log \alpha + o(1)\,.
\end{multline}
Therefore, equations \eqref{IAlphaUpper} and \eqref{IAlphaLower} show that
\begin{multline}
\label{IAlphaAsym}
I_{\alpha}(\phi,X^n) = \frac{m-1}{2}\log \frac{n}{2} + \frac{1}{2}\log\pi \\
-\log\Gamma\left(\frac{m}{2}\right) -\frac{m-1}{2(\alpha - 1)}\log \alpha + o(1)\,.
\end{multline}

We are now left with the $W_{\alpha}(\mathcal{P})$ term. Starting from \eqref{WTermDMS}, we want to find the asymptotics of the first logarithm, which is the only term dependent on $n$. From \cite{abramowitz1} we have that
\begin{equation}
\lim_{t \to\infty} t^{b-a}\frac{\Gamma(t+a)}{\Gamma(t+b)} = 1
\end{equation}
for all real numbers $a$ and $b$. Therefore, we also have
\begin{multline}
\lim_{n\to\infty} \bigg[\log\frac{\Gamma(\alpha n + \frac{m}{2})}{\Gamma(\alpha n + \frac{1}{2})} - \frac{m-1}{2}\log(\alpha n) \bigg] \\
	= \lim_{n\to\infty} \log\left[(\alpha n)^{\frac{1}{2} - \frac{m}{2}} \frac{\Gamma(\alpha n + \frac{m}{2})}{\Gamma(\alpha n + \frac{1}{2})}\right] = 0\,,
\end{multline}
or equivalently,
\begin{equation}
\log\frac{\Gamma(\alpha n + \frac{m}{2})}{\Gamma(\alpha n + \frac{1}{2})} = \frac{m-1}{2}\log(\alpha n) + o(1)\,.
\end{equation}
Plugging this into \eqref{WTermDMS} gives
\begin{equation}
W_{\alpha}(\mathcal{P}) = \frac{m-1}{2\alpha}\log(\alpha n) + \frac{1}{2\alpha} \log \pi - \frac{1}{\alpha}\Gamma\left(\frac{m}{2}\right)+ o(1)\,.
\end{equation}
Finally, plugging this and \eqref{IAlphaAsym} into \eqref{RMaxAlphaNML} leads to \eqref{RMaxAsymptotics}.

\IEEEtriggeratref{16}
\bibliographystyle{IEEEtran}
\bibliography{alphaNML}

%

\begin{IEEEbiographynophoto}{Marco Bondaschi}
is a PhD student in the Laboratory for Information in Networked Systems at \'{E}cole Polytechnique F\'{e}d\'{e}rale de Lausanne (EPFL), Switzerland. He received the Bachelor's degree in Electronics Engineering and the Master's degree in Communication Sciences and Multimedia from the University of Brescia in 2017 and 2019 respectively. His main research interests are in information theory and machine learning.
\end{IEEEbiographynophoto}

\begin{IEEEbiographynophoto}{Michael Gastpar}
received the Dipl. El.-Ing. degree from the Eidgen\"ossische Technische Hochschule (ETH), Z\"urich, Switzerland, in 1997, the M.S. degree in electrical engineering from the University of Illinois at Urbana-Champaign, Urbana, IL, USA, in 1999, and the
Doctorat \`es Science degree from the Ecole Polytechnique F\'ed\'erale (EPFL), Lausanne, Switzerland, in 2002. He was also a student in engineering and philosophy at the Universities of Edinburgh and Lausanne.

During the years 2003-2011, he was an Assistant and tenured Associate Professor in the Department of Electrical Engineering and Computer Sciences at the University of California, Berkeley. Since 2011, he has been a Professor in the School of Computer and Communication
Sciences, Ecole Polytechnique F\'ed\'erale (EPFL), Lausanne, Switzerland.
He was also a professor at Delft University of Technology, The Netherlands, and
a researcher with the Mathematics of Communications Department,
Bell Labs, Lucent Technologies, Murray Hill, NJ.
His research interests are in information theory and related coding and signal processing techniques,
with applications to learning, networks, and neuroscience.

Dr. Gastpar is a Fellow of the IEEE. He received the IEEE Communications Society and Information Theory Society Joint Paper Award in 2013 and the EPFL Best Thesis Award in 2002. He was an Information Theory Society Distinguished Lecturer (2009-2011), an Associate Editor for Shannon Theory for the IEEE TRANSACTIONS ON INFORMATION THEORY (2008-2011), and he has served as Co-Chair of the Technical Program Committee for the 2010 and 2021 International Symposia on Information Theory (Austin, TX, U.S.A., and Melbourne, Australia).
\end{IEEEbiographynophoto}
\vfill




\end{document}